\documentclass{ws-procs9x6-cpt19}
\begin{document}

\newcommand{\refeq}[1]{(\ref{#1})}
\def\etal {{\it et al.}}
%any other macros go here

\title{Recent progress on probing Lorentz violation at HUST}

\author{Ya-Fen Chen, Yu-Jie Tan, and Cheng-Gang Shao}

%\address{$^1$ University Department, University Name,\\ City, State ZIP/Zone, Country}

\address{ MOE Key Laboratory of Fundamental Physical Quantities Measurements ${\rm{\& }}$ Hubei Key Laboratory
of Gravitation and Quantum Physics, \\ PGMF and School of Physics, Huazhong University of Science
and Technology, \\ Wuhan 430074, People¡¯s Republic of China}

%\address{$^2$Group, Laboratory,\\City, State ZIP/Zone, Country}

%\author{On behalf of the LLG Collaboration}

\begin{abstract}
This work mainly introduces the recent experimental process on probing the effect of Lorentz violation (LV) at $d=6$, which is a specially striped-type structure experiment to increase the signal of LV. Besides, we also proposed a new experimental design using the striped geometry with triplex modulation to independently constrain the 14 LV coefficients with a higher sensitivity.
\end{abstract}

\bodymatter

\section{Introduction}
Gravitational phenomena is well described by General Relativity (GR), in which Einstein equivalence principle is an important foundation of it.  Since the violation of local lorentz invariance means the defect of Equivalence Principle, the study of Lorentz violation effect in the spacetime theory of gravity is a new way to explore general relativity. In the past research, the limit of LV coefficients in pure gravity are often obtained by extracting the violation signal from some short-range gravitational experimental data, such as testing the gravitational inverse-square law using a torsion pendulum \cite{2,3,4}. Recently, we have proposed a torsion scale experiment with stripe structure to test LV effect \cite{5,6}, and this experiment is ongoing. In this work, we mainly present the progress of this stripe structure experiment.

\section{LV in pure gravity sector involving quadratic couplings of Riemann curvature}
As an effective field theory, Standard Model extension (SME) provides a significant method to study the effect of LV in low-energy experiments. The pure-gravity sector can be formulated as a Lagrange density composed of the usual Einstein-Hilbert term $R$ and a cosmological constant, and the LLI violating terms expressed by an infinite series of operators with the increasing mass dimension $d$ \cite{7}, which represents LV
\begin{eqnarray}\label{equation1}
{L_d} = \frac{{\sqrt { - g} }}{{16\pi {G_N}}}(R + \Lambda + {L_M} + L_{LV}^{(4)} + L_{LV}^{(5)} + L_{LV}^{(6)} +  \cdots ),
\end{eqnarray}
here, the last three terms are the LV terms, which are constituted by the LLI violating coefficient field ${k_{\alpha \beta  \cdots }}$ and the gravitationally physical quantities. We only focus on the item of $L_{LV}^{(6)}$ to carry out our research. In spherical coordinate representation, the LV correction potential between two masses can be expressed as:
\begin{eqnarray}\label{equation2:eps}
{V_{{\text{LV}}}}(\vec r) =  - G\sum\limits_{jm} {\frac{{{m_1}{m_2}}}{{{r^3}}}} {Y_{jm}}(\theta ,\phi ){k_{jm}^{{\rm lab}}},
\end{eqnarray}
where $k_{jm}^{{\rm{lab}}}$ are the LV coefficients expressed by spherical coefficients in the lab frame. The goal of the present experiment is to give the limit of the LV coefficient.

\section{Searching LV in short-range gravitational experiments}
%In this section, we aim at introducing the experimental design and process in our work for testing LV with the short-range experiments in HUST.
From section 2, we can know the measured coefficients $k_{jm}^{{\rm{lab}}}$ are different in different frames. Therefore, through coordinate transformation, the LV coefficients at $d=6$ can be transformed to the Sun-centered frame $k_{jm}^{{\rm{N(6)}}}$. To obtained the LV coefficients from experimental data, we write the LV coefficient in the form of Fourier series expansion. Since the LV torque can be written in the form of Fourier series expansion in experiment, we can extract LV signal from the torque violation coefficients of Fourier series expansion. The corresponding torque can be expressed as:
\begin{eqnarray}\label{equation3:eps}
{\tau _{LV}} = {C_0} + \sum\limits_{m = 1}^4 {[{C_m}} \cos (m{\omega _ \oplus } T) + {S_m}\sin (m{\omega _ \oplus } T)].
\end{eqnarray}
The detailed relationship between the Fourier coefficients and LV coefficients in Equation (3) has been descriped in the reference \cite{6}.

Since the effect of LV is determined by the edge effect, we design two stripe-type experiments (horizontal and vertical stripe-type geometries) to increase this effect. Through a more in-depth analysis, we further specify that placing the experimental setup at a certain azimuth can further enlarge the LV signal. Therefore, we can use angle modulation to jointly analyze the size of the LV effect. Our most direct idea is to put the experimental device on a turntable and rotate the turntable frequency so that the limit of LV coefficients can be extracted from the triple modulation experimental data. The simplified design for new experimental setup is shown in Fig 1.

\begin{figure}
\includegraphics[width=4in]{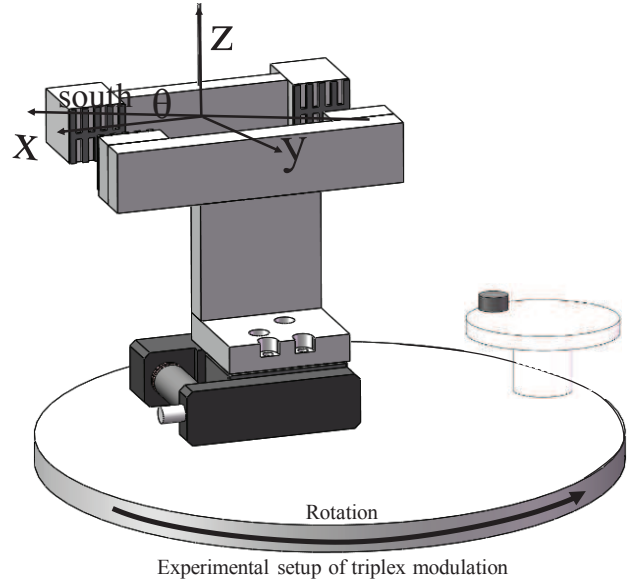}
\caption{The schematic drawing of the experimental setup for striped geometry experiments with triplex modulation.}
\label{aba:fig1}
\end{figure}

Currently, for the experiment, we successfully finished the machining, such as the tripe-shape tungsten sheet and a batch of high precision glass blocks. Figure 2 has shown tungsten in a) and glass parts in b), respectively. All of them have been measured and ensured their various geometric indices have satisfied the target accuracy requirement. Next step, we will still conduct high-precision bonding assembly.

%we continue introduce our experimental progress, but not elaborate on the processing of components here. In the process of experimental processing, to obtain materials that meet the experimental requirements, we have carried out special processing. For the whole tungsten sheet, we first use the machine to process to the similar size, then manually continue to process it to the size we need, and finally we cut the tungsten to get the stripe shape we need, like the figure 2 shown. Besides, we also need machine a batch of high precision glass blocks. Then, we measured them after processing, and the next step in our experiment is to paste the experimental parts.
%Because the size of tungsten wafer is our main error, we measured it after processing.
%In addition to the basic size measurement, we also measured the mass, gesture, parallelism and verticality of the experimental components. For very small parts, we use three coordinates to measure.

\begin{figure}
\includegraphics[width=4in]{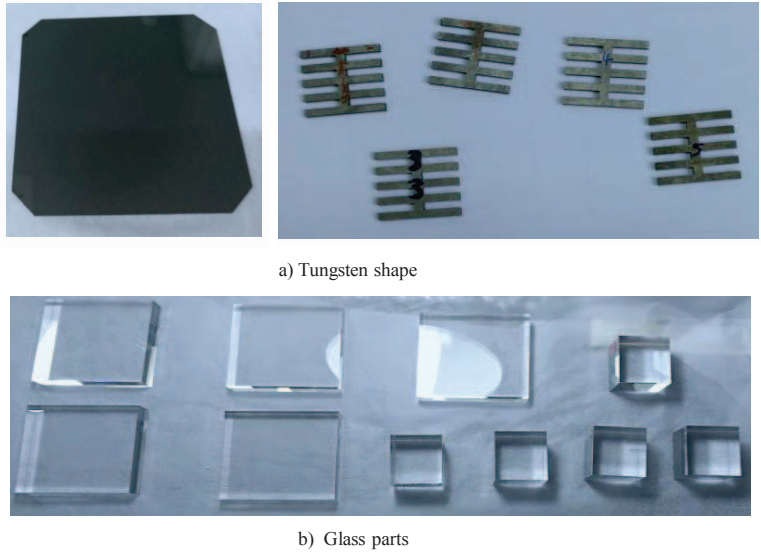}
\caption{Processed experimental parts, a) stands for the tungsten sheet and fringe structure needed in experiment, b) expresses the needed glasses in experiment.}
\label{aba:fig1}
\end{figure}

\section{Conclusion}
In this work, we mainly review our previous work on searching the LV effect in pure gravity sector in our laboratory. Based on new consideration, we propose a new experimental design aiming at searching LV signal, which increases dual modulation into triplex modulation. In this design, a rotation stage is used to rotate the whole experimental setup to modulate the LV signal, since one experiment can not give the constraints of $14$ LV coefficients independently.
%which can also avoid two or more experiments combining to constrain the $14$ LV coefficients.
%by using a combination of experiments with different angles, since one experiment can not give the constraint of 14 LV coefficients independently, which can also avoid two or more experiments combining to constrain the 14 LV coefficients.
Another important point is that our experiment completed the processing of parts, and the next step is component pasting and measurement.

\section*{Acknowledgments}
This work was supported by the National Natural Science Foundation of
China (91636221 and No. 11805074).

\end{document}